\documentclass[twocolumn,showpacs,preprintnumbers,amsmath,amssymb,prb,superscriptaddress]{revtex4}

\usepackage{epsfig}
\usepackage{dcolumn}
\usepackage{bm}
\usepackage{latexsym}

\begin{document}
\preprint{APS/123-QED}

\title{Sub-nanosecond delay of light in (Cd,Zn)Te crystals}

\author{T.~Godde}
\affiliation{Experimentelle Physik II, Technische Universit\"at Dortmund, 44221 Dortmund, Germany}
\author{I.~A. Akimov}
\email{ilja.akimov@tu-dortmund.de}
  \affiliation{Experimentelle Physik II, Technische Universit\"at Dortmund, 44221 Dortmund, Germany}
  \affiliation{A.F. Ioffe Physical-Technical Institute, Russian Academy of Sciences, 194021 St. Petersburg, Russia}
\author{D.~R. Yakovlev}
  \affiliation{Experimentelle Physik II, Technische Universit\"at Dortmund, 44221 Dortmund, Germany}
  \affiliation{A.F. Ioffe Physical-Technical Institute, Russian Academy of Sciences, 194021 St. Petersburg, Russia}
\author{H.~Mariette}
\affiliation{CEA-CNRS group "Nanophysique et Semiconducteurs", Institut N\'eel, CNRS \\
                and Universit\'e Joseph Fourier, 25 Avenue des Martyrs, 38042 Grenoble, France}
\author{M. Bayer}
\affiliation{Experimentelle Physik II, Technische Universit\"at Dortmund, 44221 Dortmund, Germany}

\date{\today}

\begin{abstract}
We study excitonic polariton relaxation and propagation in  bulk
Cd$_{0.82}$Zn$_{0.12}$Te using time-resolved photoluminescence and
time-of-flight techniques. Propagation of picosecond optical pulses
through 745~$\mu$m thick crystal results in time delays up to
350~ps, depending on the photon energy. Optical pulses with 150~fs
duration become strongly stretched. The spectral dependence of group
velocity is consistent with the dispersion of the lower excitonic
polariton branch. The lifetimes of excitonic polariton in the upper
and lower branches are 1.5 and 3 ns, respectively.
\end{abstract}




\maketitle

When the energy of a photon is close to that of an exciton resonance
in a semiconductor its group velocity for propagation may be
significantly decreased. This phenomenon originates from
exciton-photon interaction, which can be considered in terms of an
excitonic polariton (EP) quasi-particle\cite{Excitons, Pekar,
Hopfield}. The EP propagation and its dispersion have been widely
studied using various experimental techniques in different
semiconductor
materials\cite{Excitons,Ulbrich,Segawa,Itoh,Maksimiov,Frohlich,Giessen,Jutte,Shubina,
Sermage98}. Time-of-flight measurements using pulsed lasers in
conjunction with time-resolved detection resulted in group index
measurements up to several thousand in GaAs, CuCl and CdSe
crystals\cite{Ulbrich,Segawa,Itoh}. In some materials like
anthracene light may propagate even at velocities below that of
sound\cite{Maksimiov}. Additionally several coherent effects may
take place during optical pulse propagation. In linear regime
interference between the lower (LP) and upper (UP) polariton
branches leads to a beating signal, first observed for the 1S
quadrupole exciton in Cu$_2$O\cite{Frohlich}. Nonlinear effects like
self-induced transparency at the A-exciton in CdSe and at the bound
exciton in CdS have been also studied using frequency-resolved
gating \cite{Giessen} and bandwidth limited time-resolved
spectroscopy \cite{Jutte}, respectively.

Many of the experiments on EP propagation in direct band gap semiconductors like GaAs, CdSe or CdS were performed for resonance conditions on
relatively thin crystals (with thicknesses normally below tens of $\mu$m) \cite{Excitons,Ulbrich,Itoh,Giessen,Jutte}. In this case EP in both
lower and upper branches propagates through the crystal without significant losses. Although the group velocity is reduced thousand times the
optical delay is in the range of several to tens of picoseconds only. It is also known that the EP energy relaxation within the lower branch is
strongly suppressed due to the phonon bottleneck in the relaxation path \cite{Askari}. This effect may be used for long distance coherent EP
propagation and realization of long delays for optical pulses. Recently large delays of light in the order of several hundreds of picoseconds
were reported in 1~mm thick GaN crystals \cite{Shubina}. Here we report on the direct measurement of sub-nanosecond optical pulse delays in
(Cd,Zn)Te crystals which are attributed to propagation of the excitonic polariton in the lower branch. High quality samples with low impurity
concentrations result in significant transmission levels in crystals with about 1~mm thickness. The detected group velocity is only 150 times
smaller than the speed of light, but the resulting time delay accumulated in the crystal can reach almost half a nanosecond. The effects is
pronounced at low temperatures $T\leq25$~K, but disappears with temperature increase due to enhancement of the EP scattering on acoustical
phonons.

\begin{figure}
 \begin{minipage}{8.2cm}
  \epsfxsize=8.2 cm
  \centerline{\epsffile{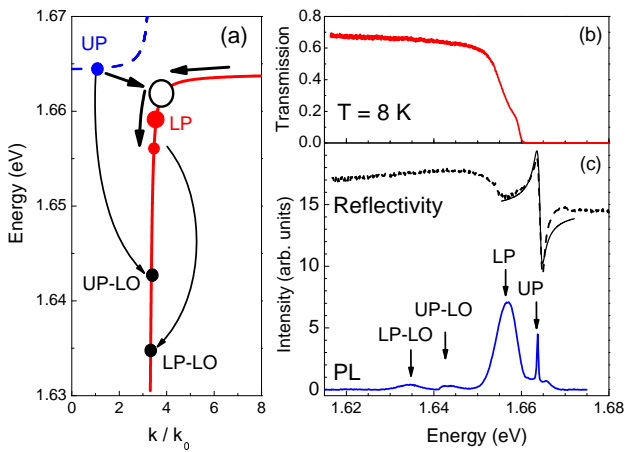}}
  \caption{\label{fig:PL}
(a) Excitonic polariton energy-wavevector dispersion ($k_0=\omega_0/c$). Circles schematically show the EP population, and arrows indicate
relaxation paths under non-resonant excitation. (b,c) Steady-state transmission, reflectivity and PL spectra at $T=8$~K. PL spectrum was taken
under non-resonant excitation $\hbar\omega_{exc}=2.33$~eV at pump power density $P=100$~mW/cm$^2$. Solid line is fit of the reflectivity
spectrum with $\hbar\omega_0=1.6638$~eV and $\Gamma=1$~meV \cite{Rfit}.}
  \end{minipage}
\end{figure}

In a single oscillator model the dependence of the dielectric
function on optical frequency $\omega$ and wavevector $k$ is given
by
\begin{equation}
\label{eq:epsilon}
    \epsilon (\omega, k) = \epsilon_B  + \frac{f \omega_0^2}{\omega_0^2-\omega^2+(\hbar k^2/M)\omega_0-i\omega\Gamma},
\end{equation}
where $\epsilon_B$ is the background dielectric constant, $f$ is the
oscillator strength, $\omega_0$ is the resonance frequency, $M$ is
the exciton effective mass and $\Gamma$ is the exciton damping
\cite{Ivchenko}. Solution of $\epsilon = (c k /\omega)^2$ yields the
dispersion relation for transversal EP modes, comprising the lower
(LP) and upper (UP) polariton branches. Here $c$ is the speed of
light in vacuum. Obviously, the single oscillator model gives a
simplified description of EP dispersion, especially in zinc-blend
crystals like (Cd,Zn)Te where the band structure is complex
\cite{Cho76}. However, being mainly interested in the LP branch at
optical frequencies $\omega\leq\omega_0$, this simple model gives
good agreement with the experimental results. The main parameters
for CdTe are known from resonant Brillouin scattering, namely
$\epsilon_B=11.2$, $f=8.8\times10^{-3}$ and $M=2.4$ in units of the
free electron mass \cite{Excitons,RBS-CdTe}. The energy dispersion
curves for the UP and LP branches are plotted in
Fig.~\ref{fig:PL}(a) for $\hbar\omega_0=1.6638$~eV, neglecting the
damping ($\Gamma=0$).

The investigated sample was prepared from the ingot of bulk Cd$_{0.88}$Zn$_{0.12}$Te crystal grown by Bridgman technique, which corresponds to a
growth process at high temperature (1200$^\circ$C). The crystal was cut along the (100) direction and after chemical-mechanical polishing had a
thickness of 745~$\mu$m. The crystal had a very good crystalline quality (rocking curve width less than 20 arcsec with an etched pits density
10$^{4}$~cm$^{-2}$). The crystal was not intentionally doped resulting in slightly p-type with a concentration in the order of
$10^{15}$~cm$^{-3}$.

For time-resolved photoluminescence (PL) and time-of-flight measurements we used a mode-locked Ti:Sa laser. The laser emitted transform limited
pulses with 1~ps or 150~fs duration at a repetition frequency of 76~MHz. The sample was mounted in a He-bath cryostat with a variable
temperature insert. The PL signal was collected in reflection or transmission geometry. For time-of-flight measurements the optical pulses hit
the sample at normal incidence and the transmitted light was detected. The signal was dispersed by a single 0.5~m spectrometer with 6.28~nm/mm
linear dispersion and detected with a streak camera. The overall temporal and spectral resolution of the experimental setup for time-resolved
measurements was about 20~ps and 1~nm, respectively. For time-integrated measurements a nitrogen cooled charge-coupled-device camera connected
to the same spectrometer with 1.3~nm/mm linear dispersion was used. Steady-state reflectivity and transmission measurements were performed using
a halogen lamp. For steady-state PL we used a continuous wave laser with $\hbar\omega_{exc}=2.33$~eV.

Figures \ref{fig:PL}(b) and \ref{fig:PL}(c) show low temperature ($T =8~K$) time-integrated steady state transmission, reflectivity and PL
spectra. From reflectivity it follows that the free exciton resonance is located at $\hbar\omega_0=$~1.6638~eV with $\Gamma=1$~meV. In the PL
spectrum the narrow peak at 1.6637~eV is due to radiative emission from the bottom of the UP branch. Consequently the PL maximum of UP is red
shifted by 0.66~meV relative to the expected value $\hbar\omega_0 + \Delta_{LT}$, where $\Delta_{LT}= f\hbar\omega_0/2\epsilon_B=0.65$~meV is
the longitudinal-transversal splitting. This may be result of EP localization on alloy fluctuations\cite{Permogorov}. The resulting energy
fluctuations are, however, small enough compared to $\Gamma$ to neglect them. At lower energies we find a broader peak centered around 1.657~eV,
which we attribute to emission from the lower EP branch. The LO phonon replica of the UP and LP lines are shifted by
$\hbar\omega_{\textrm{LO}}$=22~meV to lower energies. We observe no PL lines related to donor or acceptor bound excitons, which indicates an
exclusively pure crystal quality with low background impurity concentration.

\begin{figure}
 \begin{minipage}{8.2cm}
  \epsfxsize=8.2 cm
  \centerline{\epsffile{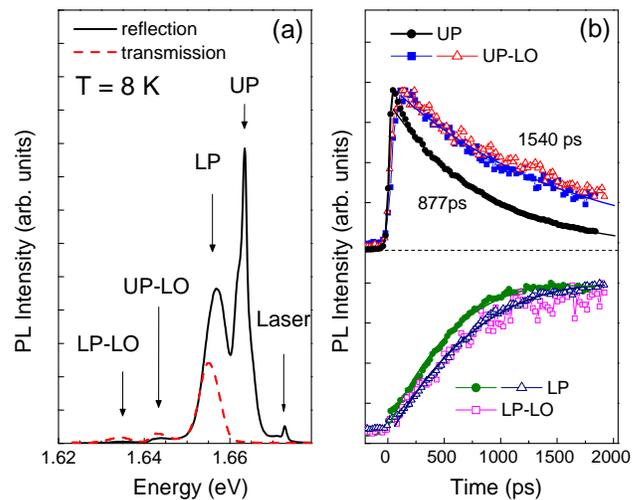}}
  \caption{\label{fig:transients}
(a) Time-integrated PL spectra under pulsed excitation with pulse duration $\tau_D$=1~ps and $\hbar\omega_{exc}= $~1.673~eV in reflection (solid
line) and transmission (dashed line) geometries. (b) Intensity transients of UP, LP and their LO phonon replicas measured in reflection (full
symbols) and transmission (open symbols). Solid lines are exponential fits. The rise time of the UP and UP-LO peaks corresponds to the decay
time of the LP peak in reflection and transmission, respectively.}
  \end{minipage}
\end{figure}

\begin{figure}
 \begin{minipage}{8.2cm}  \epsfxsize=7 cm
  \centerline{\epsffile{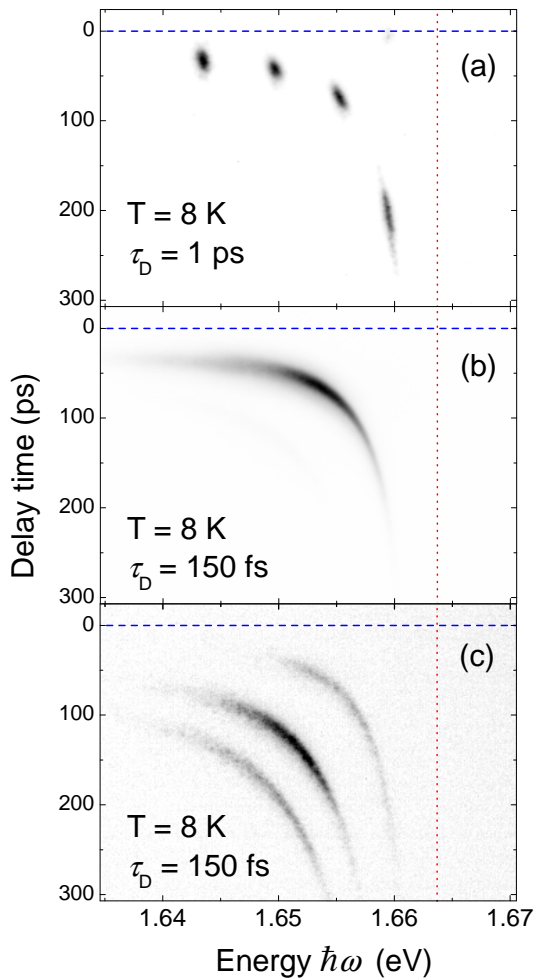}}
  \caption{\label{fig:2Dplot}
Grey-scale contour plots of the intensity of the transmitted optical pulse versus delay time and energy for pulse durations of (a) $\tau_D
=1$~ps, (b) and (c) $\tau_D =150$~fs. In (a) data for four optical pulses with different $\hbar\omega_{exc}$ are superimposed. In order to
increase the relative intensity of the replicas in (c) the sample was slightly rotated from normal incidence \cite{replicas}. Horizontal dashed
line indicates zero time, vertical dotted line gives the exciton resonance position $\hbar\omega_0 =1.6638$~eV.}
  \end{minipage}
\end{figure}

PL spectra under pulsed excitation with $\hbar\omega_{exc} = $~1.673~eV  in transmission and reflection geometries are shown in
Fig.~\ref{fig:transients}(a). It is important to keep in mind that the EP emission occurs only at the crystal surface where the conversion into
corresponding photons occurs. The UP branch line is present only in reflection geometry, while the LP peak as well as the phonon replicas of UP
and LP are clearly seen in both configurations. This is in accord with the transmission spectrum in Fig.~\ref{fig:PL}(b), i.e. the edge is
located just below $\hbar\omega_0$. This indicates that polaritons from the lower branch below the phonon bottleneck [indicated by the open
circle in Fig.~\ref{fig:PL}(a)] more likely reach the crystal surface without further scattering. Indeed, if we consider the EP energy
relaxation, which is schematically shown in Fig.~\ref{fig:PL}(a), we expect that after the excitation pulse the polaritons in the UP branch
either scatter into the LP branch via acoustic phonon emission or recombine radiatively when they are close enough to the surface. The remaining
EPs gather at the knee of the LP branch, where the probability of acoustic phonon scattering is lowest because of the phonon bottleneck
\cite{Askari}. Finally, they relax further until the group velocity becomes large enough to reach the sample boundary, where they are converted
into photons. In case of LO phonon emission the EP always reach the crystal surface and, therefore, LO replicas in the PL spectrum can be used
to monitor the corresponding EP dynamics in the sample volume \cite{Gross}.

The transients of UP, LP and their phonon replicas are shown in
Fig.~\ref{fig:transients}(b) and confirm the argumentation outlined
above. The UP peak in reflection geometry decays with a time of
877~ps, while its LO replica decays with 1540~ps, almost twice
longer. The latter corresponds to EP scattering into the LP branch.
The decay time of the UP peak may be shorter near the sample
boundary due to direct radiative recombination or additional
non-radiative recombination at the surface. Note that the UP-LO
follows the same dynamics in both reflection and transmission
geometries, monitoring the EP kinetics in the sample volume. The LP
peak is expected to have a rise time equal to the UP decay time.
This is in full agreement with the experimental data. We fit the LP
transients in reflection and transmission with double exponential
decays where the rise time corresponds to the UP and UP-LO decay,
respectively. Good agreement with experiment is obtained for a decay
time of about 3~ns, which is attributed to the EP lifetime in the
phonon bottleneck region~\cite{Askari,Cooper}. This time constant
gives the upper limit for a possible coherent delay of the optical
pulse in the LP branch.

\begin{figure}
 \begin{minipage}{8.2cm}  \epsfxsize=7 cm
  \centerline{\epsffile{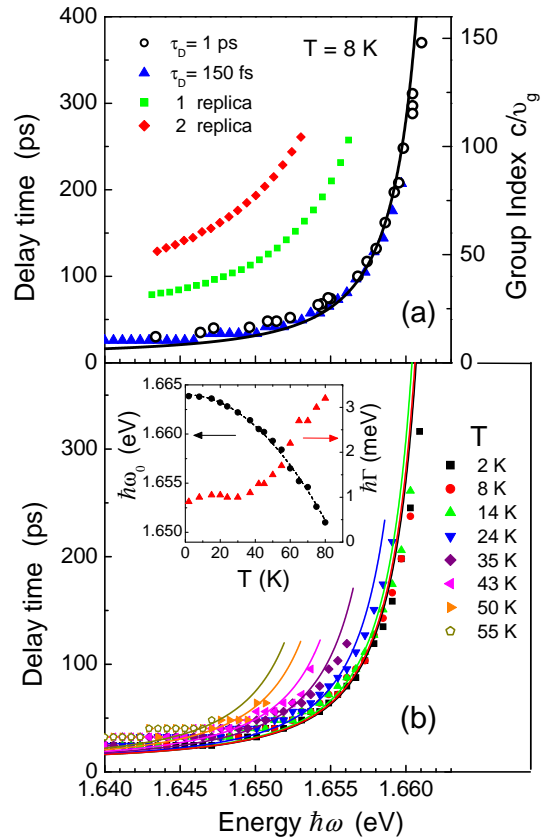}}
  \caption{\label{fig:dispersion}
(a) Delay time/group index as functions of optical pulse photon energy $\hbar\omega_{exc}$ for $\tau_D=150$~fs (solid triangles) and
$\tau_D=1$~ps (open circles). Solid squares and diamonds represent the delay of the first and second reflection replicas, evaluated from
Fig.~\ref{fig:2Dplot}(c). (b) Same as (a) with $\tau_D=150$~fs for different temperatures. The inset shows the temperature dependence of
resonance energy $\hbar\omega_0$ and exciton damping $\Gamma$, determined from the reflectivity spectra and used in the group index
calculations. Solid lines in (a) and (b) are the delay times calculated with Eqs.~(\ref{eq:epsilon}) and (\ref{eq:group}).}
  \end{minipage}
\end{figure}

After elaborating the EP kinetics we turn now to the results on optical pulse propagation in the LP branch close to the resonance frequency
$\omega\leq\omega_0$. The power density $P\leq 10$~mW/cm$^2$ was kept low enough to be in the linear regime. The contour plot of the transmitted
laser light intensity as function of delay time and photon energy is presented in Fig.~\ref{fig:2Dplot}. In case of spectrally narrow pulses
with durations of $\tau_D=1$~ps a strong dependence of pulse delay on the photon energy $\hbar\omega_{exc}$ is observed, see
Fig.~\ref{fig:2Dplot}(a). The delay time increases when $\hbar\omega_{exc}$ approaches the exciton resonance $\hbar\omega_0 =1.6638$~eV. We were
able to observe a maximum delay of 350 ps at $\hbar\omega_{exc} =1.661$ eV. The transmission of light at this photon energy decreases
drastically and is about 0.2\% only.

In case of  short laser pulses with $\tau_D=150$~fs strong distortion takes place, see Fig.~\ref{fig:2Dplot}(b). Here, the use of 12~meV
spectrally broad pulses centered at 1.653~eV allows one to monitor the EP dispersion in a single measurement. We were also able to detect the
subsequent replicas of the pulse, which originate from reflections at both crystal surfaces. The replicas can be seen especially well if the
sample is slightly tilted from normal incidence by about 5$^\circ$, see Fig.~\ref{fig:2Dplot}(c) \cite{replicas}.  As expected, the first and
second replicas appear at delays which are respectively 3 and 5 times larger than the delay of the transmitted pulse. In other words the pulses
propagate over distances of 3 and 5 times the crystal thickness of 745~$\mu$m. We note that we observe no contributions from diffusive
propagation, i.e. the photon wavevector is conserved and the pulse propagates through the sample coherently.

It is clear that the spectral dependence of the time-of-flight signal originates from strong variations of the group velocity $v_g$ near the
resonance frequency $\omega_0$. In order to quantify the experimental results we summarize the dispersion relations of the group index
$n_g=c/v_g$ and the time delay as function of the photon energy for different pulse durations $\tau_D$ and temperatures in
Fig.~\ref{fig:dispersion}. As expected, the experimental results are independent of pulse duration, i.e. the dispersion curves for $\tau_D=1$~ps
and 150~fs coincide. The temperature increase leads to a red shift of the dispersion curves. Additionally, the transmission efficiency close to
the resonance decreases strongly for temperatures above 25~K.

We calculate the group index according to
\begin{equation} \label{eq:group}
    n_g(\omega)=\frac{c}{v_g(\omega)}=\frac{d}{d\omega}\mathrm{Re}[\omega\sqrt{\epsilon_1(\omega)}],
\end{equation}
where $\epsilon_1(\omega)$ is the frequency dependent dielectric function of the LP branch satisfying Eq.~(\ref{eq:epsilon}) and the dispersion
equation for transversal modes. The UP branch is not taken into account since the photon energy is below $\hbar\omega_0$. Note that no fitting
parameters were used to reproduce the spectral dependence of the group index. As mentioned above, $\epsilon_B$, $f$ and $M$ are known, while
$\omega_0$ and $\Gamma$ are directly taken from the reflectivity measurements at different temperatures, see inset in
Fig.~\ref{fig:dispersion}(b). The results of calculations are shown by the solid lines in Fig.~\ref{fig:dispersion} and are in good agreement
with the experimental data. The temperature increase leads not only to a red shift of the energy gap, in correspondence with Varshni's law, but
also to an increase of damping $\Gamma$, as is clearly seen from the inset of Fig.~\ref{fig:dispersion}(b). The temperature growth of $\Gamma$
is associated with enhanced EP scattering on acoustical phonons.

In conclusion, we directly observed delays of light pulses up to 350~ps in a 745~$\mu$m thick (Cd,Zn)Te crystal. The experimental data are well
reproduced using the dispersion relation for the lower EP branch calculated in a single oscillator model. From time-resolved measurements we
evaluate lifetimes of EPs in the upper and lower branches of 1.5 and 3~ns, respectively. The realization of large optical delays in ternary
alloy crystals like (Cd,Zn)Te is especially attractive as these systems allow control of the resonant frequency over a broad spectral range from
1.6 to 2.4~eV by tailoring the Zn and Cd contents.

The authors are grateful to D.~Fr\"ohlich, A.N.~Reznitsky and M.M.~Glazov for useful discussions.


\begin{references}
\bibitem{Rfit} $R=\left|(1-\bar{n})/(1+\bar{n})\right|^2$, $\bar{n}=(\sqrt{\epsilon_1\epsilon_2}+\epsilon_B)/(\sqrt{\epsilon_1}+\sqrt{\epsilon_2})$,
where $\epsilon_1(\omega)$ and $\epsilon_2(\omega)$ are the
frequency dependencies of the dielectric functions corresponding to
the LP and UP branches, respectively \cite{Ivchenko}.
\end{references}
\end{document}